# Long-range interaction effects on the phase transition, mechanical effect, and electric field response of BaTiO$_3$ by machine learning potentials

Po-Yen Chen[a)], Teruyasu Mizoguchi[a),b)].
a) Department of Materials Engineering, the University of Tokyo, Tokyo, Japan
b) Institute of Industrial Science, the University of Tokyo, Tokyo, Japan.
Corresponding authors and emails
Po-Yen Chen: poyen@iis.u-tokyo.ac.jp,
Teruyasu Mizoguchi: teru@iis.u-tokyo.ac.jp

**Abstract:** Bulk materials are governed by both short-range and long-range interactions, both of which are naturally captured in conventional density functional theory (DFT) calculations through Ewald summation of electrostatic contributions. In contrast, machine learning potentials (MLPs) typically rely on local atomic environment descriptors, and long-range interactions are often neglected. Such approximations may introduce systematic energetic errors and lead to inaccuracies in predicted material properties. To systematically investigate the impact of long-range interactions in ferroelectric BaTiO$_3$ within the framework of MLPs, we developed a long-range MACELES model and compared its performance with the previously reported BaTiO$_3$ MACE model across four key properties (phonon dispersion, phase transition behavior, mechanical response, and ferroelectric properties including dielectric constants). We find that qualitative behaviors, including phase transitions, stress-induced polarization switching, and polarization–electric field hysteresis, are consistently reproduced by both models. In contrast, quantitative properties such as transition temperatures, elastic constants, and dielectric constants exhibit systematic improvements in MACELES model, highlighting the importance of incorporating long-range electrostatics for accurately describing the structural and dielectric responses of BaTiO$_3$. These results suggest that while long-range interactions play a role in improving quantitative accuracy, their omission does not significantly alter the qualitative ferroelectric behavior of BaTiO$_3$.

**Article Highlights**
- Role of long-range electrostatics in machine-learning potentials for ferroelectric BaTiO$_3$ is clarified.
- Phase transition behavior and polarization switching remain largely unchanged.
- Transition temperature and dielectric response are sensitive to long-range interactions.

## 1. Introduction

With the rapid advancement of machine learning (ML) techniques, computational methodologies for materials investigation have undergone substantial transformation. In particular, for atomic-scale molecular dynamics (MD) simulations, machine learning potentials (MLPs) have emerged as highly accurate and computationally efficient alternatives to ab initio molecular dynamics (AIMD)[1]. MLPs can achieve near density functional theory (DFT) accuracy while several orders of magnitude being more computationally efficient than AIMD, thereby enabling simulations over significantly larger length and time scales. In DFT calculations, long-range electrostatic interactions are rigorously treated through Ewald summation under periodic boundary conditions, ensuring that Coulomb interactions are accurately captured across all length scales. Despite these advantages of MLPs in computational efficiency, most MLP frameworks, including the widely used MACE model[2], decompose the total energy into atomic site energies defined solely within a finite cutoff radius [3-6]. This locality assumption, while computationally convenient, means that phenomena governed by long-range electrostatics, such as LO-TO splitting in phonon dispersion, polarization-induced electric fields, and dipole-dipole interactions extending beyond the cutoff, are not explicitly encoded in the model. Consequently, such approximations may introduce systematic energetic errors and potentially lead to inaccuracies in predicted material properties.

Despite these concerns, previous studies have suggested that, for certain ferroelectric systems such as BaTiO$_3$, a prototypical perovskite ferroelectric widely used as a model system for investigating polarization dynamics and phase transitions, neglecting long-range electrostatic interactions may not significantly affect MD simulations or phase transition behavior[7, 8]. However, the energetic consequences of neglecting long-range electrostatic interactions remain largely unexplored in the context of MLPs. In particular, inadequate treatment of long-range electrostatics may introduce systematic energy deviations, which could potentially lead to spurious metastable phases or other unphysical artifacts in atomistic simulations[3, 8]. Therefore, a systematic and direct comparison between short-range and long-range MLP frameworks is essential to quantitatively assess the role of long-range electrostatics in ferroelectric MD simulations.

To investigate the long range interaction, various strategies have been proposed to combine electrostatics with MLPs, including using global geometrical information, encoding electronic information, or targeting the response properties[9]. Recently, Kim et al. proposed a novel machine learning framework known as the latent Ewald summation (LES) model[10-

[13], which can be viewed as a hybrid of the three aforementioned approaches. In this approach, latent atomic charges are inferred directly from structural information and incorporated into an Ewald summation scheme to capture long-range electrostatic interactions. This framework enables an implicit and data-driven treatment of long-range electrostatics without requiring predefined atomic charges. As the LES framework has been integrated into the MACE architecture, the present work aims to systematically investigate the impact of long-range interactions in ferroelectric MD simulations by comparing two MACE-based models: a conventional short-range MACE model and a MACE model augmented with LES (hereafter referred to as MACELES). To this end, we perform a systematic investigation of long-range interaction in phonon dispersion, phase transition behavior, mechanical effect, and electric-field responses.

## 2. Experimental work

### 2.1 MACELES model construction and validation

In this study, to isolate the effect of the long-range interaction model, the same training database used in the previous MACE model[14] was adopted for constructing the MACELES potential. The dataset consists of 4,045 AIMD configurations of $BaTiO_3$ obtained from a 2×2×2 supercell across the four structural phases (rhombohedral, orthorhombic, tetragonal, and cubic) over a wide temperature range from 1 to 4000 K.

For model training, 95% of the dataset was used as the training set, while the remaining 5% was reserved for validation. The model was trained with a batch size of 10 and a learning rate of 0.01 Å. A patience of 20 epochs was used for the early stopping criterion. Prior to reaching the patience threshold, the loss was optimized using an exponential moving average (EMA). Once the threshold was reached, the training procedure switched to stochastic weight averaging (SWA) and continued until a total of 1000 epochs were completed. The energy and force contributions to the loss function were set to the default values used in the MACE framework. The AMSGrad optimizer was employed to improve convergence stability during training.

To evaluate the prediction ability of MACELES model, we focused on the solid phases of the test database in previous studies[14], which excludes the high-temperature and defect-containing structures, and the energy and force comparison of MACELES and MACE is shown in Figure S1a and S1b. The MAEs of energy and force calculation are 1.01 meV/atom and 11.6 meV/Å, respectively, which is quite similar to the MACE model of 0.85 meV/atom and 17.5 meV/Å, indicting the similar prediction ability of these two models.

### 2.2 Phonon dispersion comparison

The phonon dispersions of the tetragonal phase of BTO was calculated using DFPT and ASE combined with MLFF model, and the initial tetragonal structure were optimized using DFT with the PBEsol exchange-correlation functional. For the DFT part, we utilized DFPT to calculate phonon dispersion on a 3x3x3 supercell, using a plane wave cutoff energy of 500 eV and convergence criterion of $10^{-8}$ eV for total energy. Besides, we also calculated the Born effective charges for the primitive tetragonal $BaTiO_3$ cell to conduct the non-analytic correction (NAC), and the cutoff energy is 500 eV and convergence criteria is $10^{-5}$ eV. For the MACE- and MACELES-based calculations, phonon dispersion was calculated obtained using the finite displacement method[15] with the supercell size from 3x3x3 to 6x6x6, and forces were calculated using the trained MACE and MACELES models, and the dynamical matrices at arbitrary q-points were constructed by Fourier transforming the real-space interatomic force constants.

### 2.3 Hysteresis loop simulation

In the hysteresis loop simulation, we utilize MLP-MD to conduct the P-E hysteresis loop simulation. As mentioned in our previous study, we achieve the electric field response during MLP-MD simulation by implementing an additional machine learning model, Equivar_eval model[16, 17], which can calculate the born effective charge directly from the structural information. The Equivar_eval model we utilized in this model is the same as previous study, which utilizing the BEC data calculated by DFPT method based on our BTO database to finetune and provide highly accurate results with MAE of 0.0697 e.

Based on our previous work, we can compute the external electric field via the formula $F_{ext} = |e|\mathcal{E}_\beta Z^*_{\kappa,\beta\alpha}$, where $e$ is the elementary charge, $\mathcal{E}$ is the electric field, $Z$ is the BEC tensor. Consequently, the total force can be calculated by $F_{total} = F_{MLP} + F_{ext}$, where $F_{MLP}$ is the force calculated by machine learning potential. To avoid the effect of the electric field scanning rate, or scanning frequency, all of the hysteresis loop simulation was utilized a triangular-wave electric field with $\mathcal{E}_{max}$ = 100 kV/cm, the frequency is 2.5 GHz, and the temperature is maintained at 250 K.

## 3. Results and discussion

*3.1 Phonon dispersion*

Before investigating the material properties, it is necessary to verify that long-range interactions are properly captured in the MACELES model. Phonon dispersion provides a direct probe for this purpose, as long-range Coulomb interactions give rise to the longitudinal optical–transverse optical (LO–TO) splitting near the Γ point. Therefore, the presence of LO–TO splitting in the calculated phonon spectrum can serve as an indicator of whether long-range electrostatic interactions are correctly described by the machine learning potential. To examine this effect, tetragonal $BaTiO_3$ supercells with sizes ranging from 3×3×3 to 6×6×6 were constructed, and the corresponding phonon dispersions were calculated. Figures 1a–1d present the phonon dispersion obtained using the MACE and MACELES models for different supercell sizes by finite-displacement calculation, together with reference results from density functional perturbation theory (DFPT)[18, 19] calculations with and without the non-analytic correction (NAC). In polar materials such as $BaTiO_3$, long-wavelength LO phonons generate macroscopic electric fields through ionic displacements, which in turn act back on the ionic motion via long-range Coulomb interactions. These long-range contributions are not captured by the short-range interatomic force constants alone, and must be treated separately through the NAC, which analytically incorporates long-range electrostatic interactions into the dynamical matrix using the Born effective charge (BEC) tensor and the high-frequency dielectric constant obtained from DFPT. The DFPT results with NAC capture the LO–TO splitting characteristic of tetragonal $BaTiO_3$, whereas the results without NAC represent only the short-range interaction contribution. This distinction makes the comparison particularly meaningful in the present context: the DFPT without NAC serves as a natural reference for the short-range MACE model, while the DFPT with NAC corresponds to the physical regime that MACELES aims to reproduce through its latent Ewald summation framework.

As shown in Figure 1, the overall phonon spectra predicted by both MACE and MACELES are generally consistent with the DFPT results. With increasing supercell size, the phonon dispersion obtained from both models gradually approaches the DFPT results with NAC. However, a notable difference emerges near the Γ point, most clearly visible by comparing the red (MACELES) and orange (MACE) curves in the Γ–Z region across panels (a) to (d): only the MACELES model exhibits a gradual shift of the LO branch toward the Γ point as the supercell size increases from 3×3×3 to 6×6×6, progressively approaching the DFPT result with NAC (blue dashed line). In contrast, the MACE model (orange) shows no such size-dependent shift of the LO branch, remaining close to the DFPT result without NAC (green dashed line) regardless of supercell size. This behavior reflects the finite-size convergence of the LO–TO splitting, which is known to converge slowly with respect to supercell size due to the long-range nature of the dipole–dipole interaction[20]. The convergence is particularly evident in panel (d) (6×6×6), where the MACELES LO branch near the Γ point approaches the DFPT with NAC reference, while the gap between MACE and the NAC reference remains essentially unchanged. Similar size-dependent convergence behavior has previously been reported by Zhang et al. for bulk NaCl using a long-range model[20]. Therefore, the phonon dispersion analysis confirms that the MACELES model is capable of capturing the long-range electrostatic interactions in $BaTiO_3$ systems.

Furthermore, an additional signature of long-range electrostatic interactions can be observed in the Γ–X region at approximately 20 THz. In the MACELES results (red curves), a sharp upward shift of the LO branch near the Γ point, reflecting the LO–TO splitting characteristic of polar materials with long-range electrostatic interactions, as seen in the DFPT results with NAC (blue dashed line). However, the MACELES LO branch appears as a sharper, more spike-like feature compared to the smoother rise in the DFPT with NAC reference, suggesting that while MACELES qualitatively captures the long-range electrostatic character, the quantitative reproduction of the LO branch shape near the Γ point remains imperfect. The spike originates from the Gibbs oscillation. In the finite-displacement phonon calculations based on the MACE and MACELES models, the Fourier interpolation of real-space force constants cannot capture the non-analytic behavior in the limit $q \to 0$. As a result, a Gibbs oscillation emerges instead of the LO-TO splitting associated with long-range dipole-dipole interactions[20, 21]. Nevertheless, this feature is entirely absent in the MACE model (orange), confirming that it is a direct consequence of the long-range interactions incorporated in MACELES.

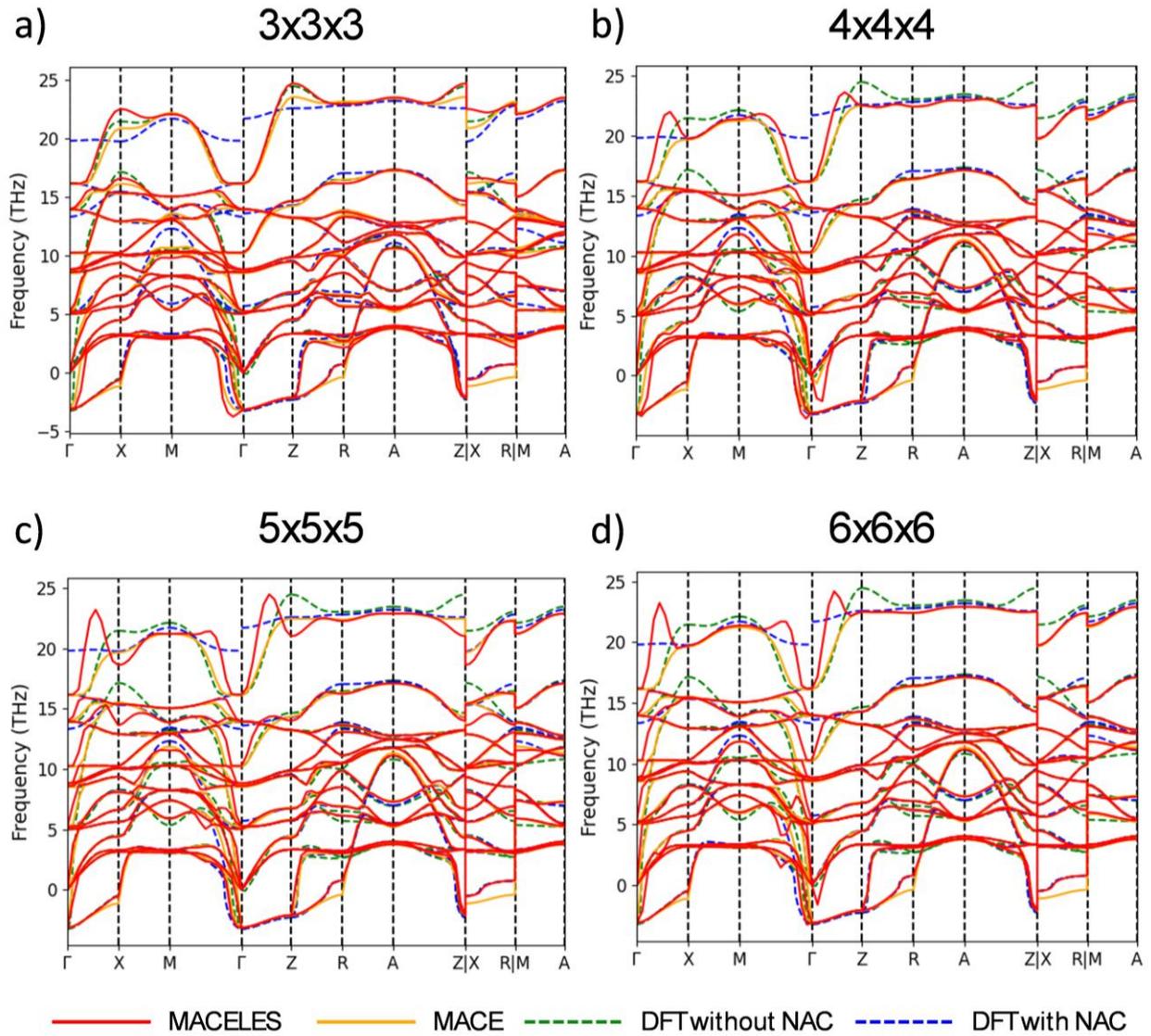

Figure 1 The comparison of phonon dispersion between MACELES (red), MACE model (orange), DFT without non-analytic correction (NAC) (green dashed), and DFT with NAC (blue dashed) for the supercell size from 3×3×3 to 6×6×6 for tetragonal $BaTiO_3$.

*3.2 Phase transition temperature*

In our previous study on $BaTiO_3$[14], the phase transition sequence from rhombohedral (R) to orthorhombic (O), tetragonal (T), and finally cubic (C) phases was successfully reproduced using a fine-tuned MACE model. The predicted transition sequence is consistent with experimental observations and previous simulation studies[7, 22-24]. To investigate the effect of long-range interactions on the phase transition behavior, MD simulations using the MACELES model were performed under the same heating protocol as in our previous work, where the system was gradually heated from 1 to 350 K at a constant rate of 2.5 K/ps by NPT ensemble. The temperature-dependent lattice constants are shown in Figure 2a. As shown in Figure 2a, the temperature evolution of the lattice constants obtained from MACELES exhibits the same qualitative trend as that predicted by MACE. This indicates that both models reproduce the same phase transition sequence (R–O–T–C), suggesting that the inclusion of long-range interactions does not significantly alter the transition pathway. This observation is consistent with previous Gigli et al.'s suggestions[7] and the effective atomic potential results reported in Yu et al[8].

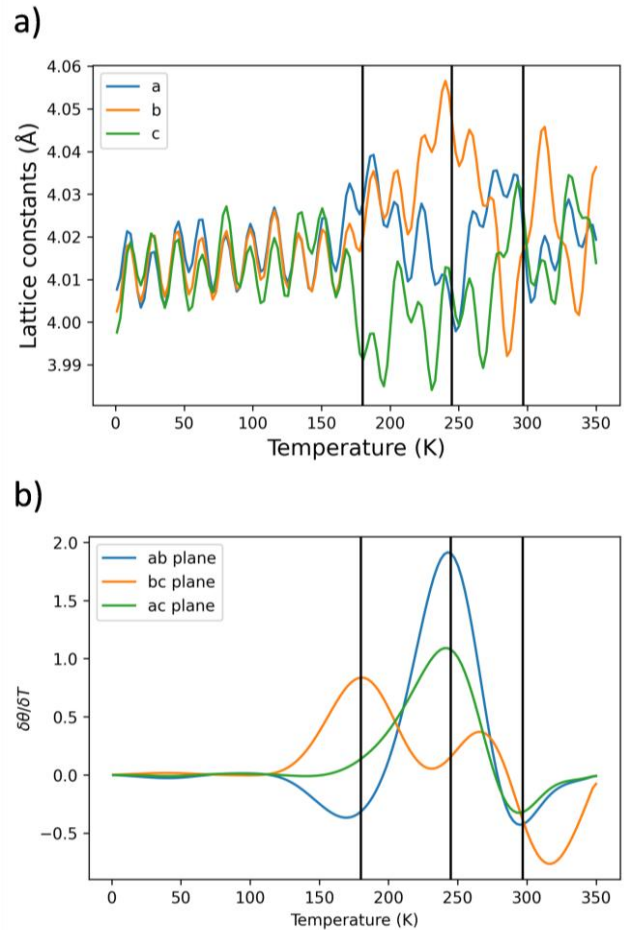

Figure 2 Temperature-dependent a) lattice constants (a-axis: blue, b-axis: orange, c-axis: green) and b) differentiated angles of Ti displacement vectors projected onto three crystallographic plane (ab-plane : blue, bc-plane : orange, and ac-plane : green) and denoised using Gaussian filter from 1 to 350 K.

Since the temperature-dependent lattice constants have quite large vibration, to capture the more precise transition temperature, same as previous study[14], we utilized the temperature-dependent differentiated angles of Ti displacement after denoising with a Gaussian filter[25], where the peak indicates the sudden change in Ti displacement and corresponds to transition temperature. Whereas, we found that the phase transition temperatures, R→O, O→T, T→C, predicted by MACELES (180, 245, 297 K),respectively, are slightly higher than those obtained from MACE (150, 225, and 290 K)[14]. To understand the origin of this difference, we further examine the supercell-size dependence of the fully relaxed unit-cell volume for both MACE and MACELES models with the maximum relaxation force of 0.05 eV/Å by MDMin algorithm in ASE package, as shown in Figure 3. The results reveal that, for both models, the unit-cell volume exhibits a sudden increase when the supercell size reaches 6×6×6. Previous studies have reported that the underestimation of the unit-cell volume by the PBEsol exchange–correlation functional[26] leads to an underestimation of the phase transition temperature[7,14]. Therefore, the slightly larger unit-cell volume predicted by the MACELES model can explain the observed increase in transition temperature compared to the MACE model although the effect is relative tiny compared with the effect on the XC functionals. A similar tendency associated with long-range electrostatic interactions has also been reported in Yu et al.'s studies, although it has not yet been systematically investigated[8].

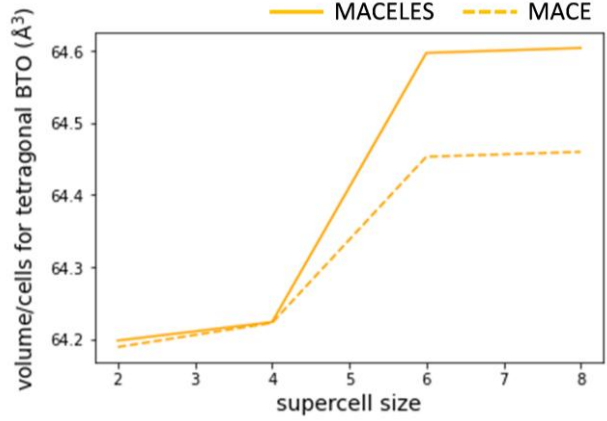

Figure 3 Supercell size-dependent unit cell volume for the relaxed tetragonal BaTiO$_3$ by MACELES (solid line) and MACE (dashed line) model.

As for the origin of the increased unit-cell volume observed in larger supercells, we attribute this behavior to a finite-size effect. As shown in the phonon dispersion in Figure 1, both the MACE and MACELES models exhibit imaginary phonon modes, indicating that the current <001> polarization configuration is not the most stable state of the system. Experimental studies[27] have suggested that in the tetragonal phase of BaTiO$_3$, the local polarization direction can deviate from the $c$-axis by approximately 43°, which is commonly interpreted within the order–disorder model of ferroelectricity[28,29]. In small supercells, the periodic boundary condition strongly constrains the polarization direction, stabilizing the ideal <001> orientation. In contrast, larger supercells impose fewer constraints, allowing the system to relax toward a more stable configuration with a rotated polarization direction accompanied by slightly structural distortion. As illustrated in Figure S2a, when the supercell size reaches 6×6×6, the lattice constant $c$ slightly decreases while the lattice constants $a$ and b increase. To further verify this interpretation, the relaxed structures of 4×4×4 and 8×8×8 supercells obtained from the MACE and MACELES models were examined, as shown in Figures S2b–S2e. Both models exhibit nearly perfect <001> polarization in the 4×4×4 supercell, whereas noticeable polarization deviation appears in the 8×8×8 supercell. The angle between the polarization direction and the $c$-axis is approximately 34.5° in the MACE model and 38.9° in the MACELES model. This difference in polarization rotation, together with the accompanying structural distortion, is likely responsible for the slight difference in unit-cell volume between the two models. These results suggest that long-range interactions mainly affect the quantitative magnitude of the structural distortion, while the overall qualitative behavior remains consistent.

### 3.3 Mechanical effect

Next, we investigate the mechanical stress effect on BaTiO$_3$ with and without long-range interactions. To calculate the elastic constants, we utilized the slightly strained structures with -0.5% and +0.5% deformation and extracted by fitting the strain-energy relationship to a second-order polynomial. As a first step, the elastic constants of the unit cell were calculated for the MACELES model and compared with those obtained from the MACE model in our previous study[14], as well as with GGA-PBEsol[30] calculations and available experimental data[31-36], as summarized in Table 1. The results indicate that the elastic constants predicted by the MACELES model are slightly lower than those obtained from the MACE model and are closer to the values calculated using the GGA-PBEsol functionals. Although both MACE and MACELES predictions are similar to experimental data, the reduced elastic constants in the MACELES model suggest that the lattice is mechanically softer than that predicted by the MACE model. This softer lattice response is also consistent with the larger structural distortion and polarization rotation observed in larger supercells, as discussed in Section 2.2.

Table 1 Comparison of the elastic constants obtained by MACELES, MACE model, GGA-PBEsol, and experimental data.

|  | $C_{11}$ | $C_{12}$ | $C_{13}$ | $C_{33}$ | $C_{44}$ | $C_{66}$ |
| --- | --- | --- | --- | --- | --- | --- |
| MACELES | 286 | 98 | 89 | 128 | 111 | 121 |
| MACE[14] | 321 | 125 | 113 | 164 | 115 | 125 |
| GGA-PBEsol[30] | 281 | 98 | 89 | 128 | 111 | 121 |
| Experiment[31-36] | 211~275 | 107~179 | 104~151 | 126~165 | 42~64 | 93~134 |

To further examine the mechanical response, stress-dependent lattice constants were calculated for both the MACE and MACELES models under applied stresses ranging from 0 to 160 MPa with an interval of 40 MPa and maintained at 250 K for 20 ps. The results are shown in Figure 4 for MACELES model, and the black line shows the coercive stress obtained by previous MACE study[37]. For both models, a sharp change in the lattice constants is observed at approximately 120 MPa, indicating the occurrence of stress-induced polarization switching. The corresponding coercive stress is therefore approximately 120 MPa for both models, showing a similar value with experimental data[38]. This result suggests that the inclusion of long-range interactions has a neglectable effect on the coercive stress in $BaTiO_3$ within the present simulation conditions.

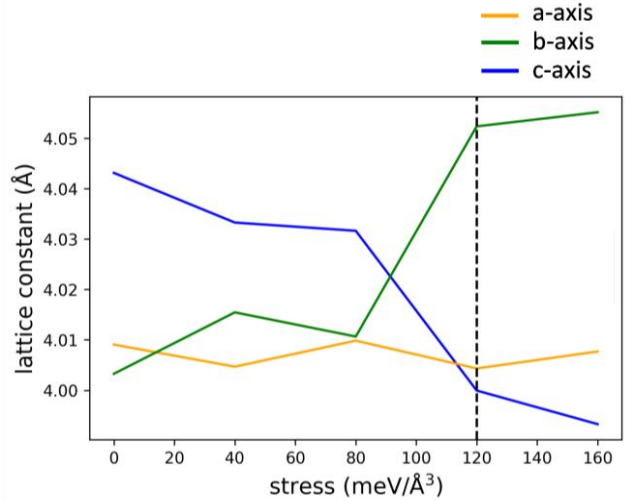

Figure 4 Stress-dependent lattice constants (a-axis: orange, b-axis: green, and c-axis: blue) at 250K simulated by MACELES. The black dashed line is the coercive stress investigated by our previous study by MACE model[14].

*3.4 Ferroelectric properties*

In the final part, we investigate the effect of long-range interactions on the prediction of ferroelectric properties using the MACE and MACELES models. To evaluate the ferroelectric response, we employed the molecular dynamics framework developed in our previous work[37, 39], which integrates the MLP model with a Born effective charge (BEC) prediction model, the Equivar model[16, 17], to simulate the polarization–electric field (P–E) hysteresis loop. Following our previous study[37], a triangular electric field was applied with the same scanning rate (2.5 K/ps) to $BaTiO_3$ at 250 K under the NPT ensemble, as shown in Figure 5a. The results show that both the MACE and MACELES models produce similar ferroelectric hysteresis behavior, with comparable remnant polarization and coercive field values. Besides, we can also observed the sharp pulse at a/b orientation in Figure 5b when the polarization switching happened, indicating the two-step polarization switching happened. This is consistent with the observation in our previous study and other simulation reports. This indicates that the inclusion of long-range interactions has a limited influence on the overall hysteresis characteristics of $BaTiO_3$, which is similar to the observations reported in reported in Yu et al with 50K[8].

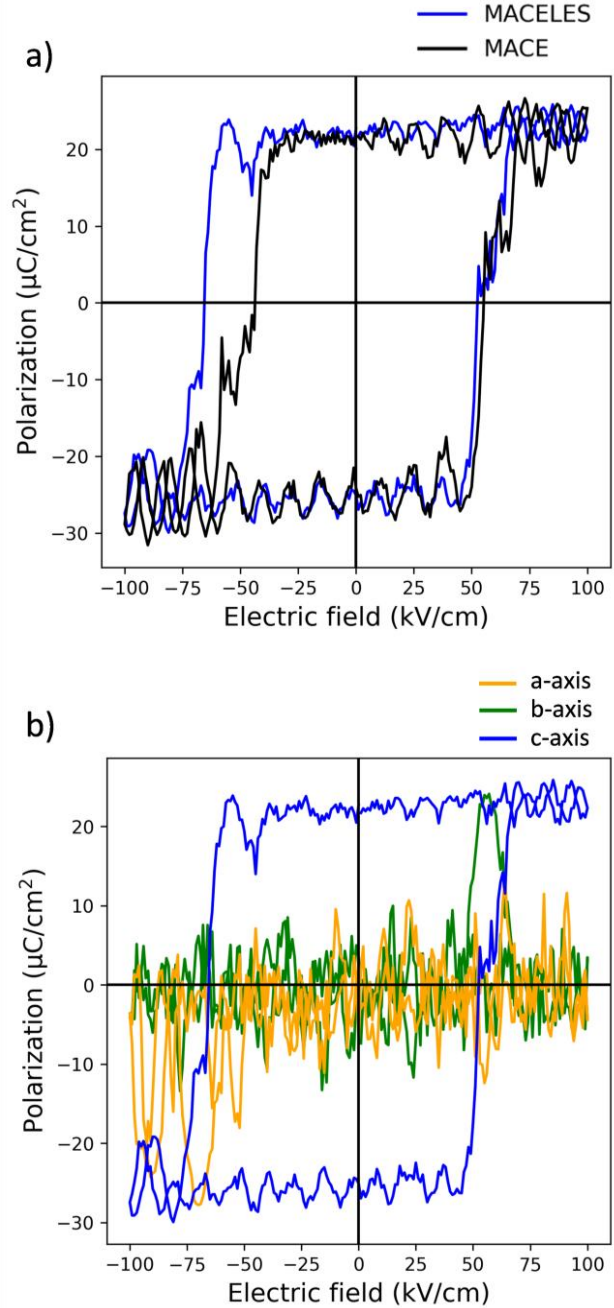

Figure 5 a) the comparison of hysteresis loop of MACELES (blue) and MACE (black) model[37] of the tetragonal 8x8x8 BaTiO$_3$ supercell, and b) the polarization change of three axes (a-axis: orange, b-axis: green, and c-axis: blue) during the hysteresis loop.

In addition to the hysteresis behavior, we further evaluated the dielectric constants predicted by the two models. In our previous MACE simulations, the out-of-plane dielectric constant ($\varepsilon_c$) was found to be close to the experimental value, whereas the in-plane dielectric constant ($\varepsilon_a$) was underestimated, being approximately one-third smaller than the experimental value. To examine the influence of long-range interactions on the dielectric response, we calculated $\varepsilon_c$ from the slope of the P–E hysteresis loop shown in Figure 5, and $\varepsilon_a$ from the linear polarization response under a small electric field range from −20 to 20 kV/cm with an interval of 10 kV/cm, where each field was maintained for 20 ps as shown in Figure S3. The obtained values of $\varepsilon_c$ and $\varepsilon_a$ are 171 and 1440, respectively. Compared with the values of 153 and 1070 obtained by MACE model, the $\varepsilon_a$ becomes slightly larger when long-range interactions are included. This behavior can be related to the tetragonality of BaTiO$_3$ Previous study reported by Sahashi et al. have shown that the tetragonality strongly influences the dielectric constant, and for the value of c/a is smaller the $\varepsilon_a$ tends to become larger[40]. In our simulations, as discussed in Section 2.2, the MACELES model exhibits a slightly reduced tetragonality due to the softer lattice response, and the phenomenon in MACELES is stronger than MACE model, and the value of c/a in 8x8x8 tetragonal BaTiO$_3$ obtained by MACELES is 1.018, which is smaller than the value of 1.023 that obtained by MACE model. The reduced tetragonality weakens the polarization anisotropy and allows larger in-plane polarization fluctuations, which leads to a slightly enhanced in-plane dielectric constant $\varepsilon_a$, while both of them is still lower than the experimental measuremnt[41].

*3.4 Discussion: Qualitative topology versus quantitative curvature of the potential energy surface*

The results presented in Sections 2.1-2.4 reveal a consistent and physically meaningful pattern across all investigated properties: the inclusion of long-range electrostatic interactions via the LES framework leaves qualitative ferroelectric behaviors essentially unchanged, while producing moderate but systematic modifications in quantitative predictions. We argue that this separation is not coincidental but reflects a fundamental distinction between two complementary aspects of the potential energy surface (PES): its topological structure and its local curvature and the precise positions of its minima.

In many systems, such as doped or layered materials, neglecting long-range electrostatic interactions can significantly influence dynamical behavior[42, 43]. However, in perovskite ferroelectrics such as $BaTiO_3$, the structural phase transitions and ferroelectric properties are primarily driven by the ferroelectric instability arising from the competition between short-range interactions and dipole–dipole interactions[8, 44]. Importantly, a significant portion of this interaction is already contained in the interatomic force constants at short distances, extending to the first-neighbor shells. Consequently, qualitative behaviors, such as the phase transition sequence (R-O-T-C), the occurrence of stress-induced polarization switching, and the overall shape of the polarization-electric field hysteresis loop, are predominantly determined by short-range interatomic interactions, including covalent bonding character, steric repulsion, and nearest-neighbor electrostatics, all of which are well captured within the local cutoff radius of the MACE framework. Long-range electrostatic interactions, while present, do not fundamentally alter which phases are stable or the sequence in which transitions occur. This interpretation is consistent with prior suggestions by Gigli et al.[7] and Yu et al.[8] that short-range models are sufficient for capturing the qualitative phase behavior of $BaTiO_3$, and provides a physical basis for understanding why.

In contrast, quantitative properties are sensitive to the finer details of the PES. Phonon frequencies, elastic constants, and dielectric constants are all directly related to the second derivatives of the potential energy, with respect to atomic displacements, strain, and polar distortion modes, respectively, and thus directly reflect the local curvature of the PES. This effect is also reflected in the energy profile along representative distortion paths, where neglecting long-range electrostatic interactions leads to different energy variations along the same pathway[7].

Phase transition temperatures depend on both the relative depths of the energy minima and the vibrational entropy encoded in the curvature of each well. Long-range electrostatic interactions contribute measurably to both: in MACELES, the equilibrium unit-cell volume is slightly enlarged and the effective force constants are softened, which directly shifts the transition temperatures, elastic constants, and dielectric constants relative to MACE. The in-plane dielectric constant $\varepsilon_a$ shows the most pronounced response, with a ~35% increase, reflecting its direct sensitivity to PES curvature along polar distortion modes.

The phonon dispersion provides additional support for this picture. Since phonon frequencies are second derivatives of the potential energy with respect to atomic displacements, they are direct probes of PES curvature. The LO-TO splitting near the Γ point, present in MACELES but absent in MACE, confirms that the LES framework modifies the PES curvature through long-range electrostatics. The remaining quantitative discrepancy with the DFPT with NAC reference, however, reflects a fundamental difference: DFPT determines LO-TO splitting from Born effective charges computed via linear-response theory, whereas LES infers latent charges implicitly from energy and force data alone. The latent charges therefore qualitatively mimic BECs but do not reproduce their magnitude exactly, a limitation that BEC-supervised training could potentially address.

Taken together, the results suggest a practical guideline: short-range MLPs are sufficient for qualitative exploration of phase behavior and switching phenomena, while long-range models are necessary for quantitative accuracy in vibrational, mechanical, and dielectric properties. This distinction originates from the nature of the PES: long-range electrostatic interactions change its curvature and shift the energy minima, but do not alter which phases are stable. Therefore, whether to include long-range interactions in MLPs should be decided not by the choice of material, but by which property needs to be accurately predicted.

## 4. Conclusions

In conclusion, we systematically investigated the effect of long-range interactions on the ferroelectric properties of $BaTiO_3$ by comparing the material properties predicted by the MACE and MACELES models. The phonon dispersion analysis confirms that the MACELES model successfully captures long-range electrostatic interactions, as evidenced by the emergence of the characteristic LO–TO splitting behavior. For the phase transition behavior, the MACELES model reproduces the same phase transition sequence as the MACE model, while predicting slightly higher transition temperatures. This difference is attributed to the small increase in the equilibrium unit-cell volume observed in the MACELES simulations. Regarding the mechanical response, both models predict similar coercive stress values. However, the elastic constants obtained from the MACELES model are slightly smaller than those from the MACE model and are closer to the values calculated using the GGA-PBEsol functional, while both remain within reasonable agreement with experimental data. For the ferroelectric response, both models produce similar polarization–electric field hysteresis loops with comparable remnant polarization and coercive field values. A slight change in the dielectric constant is observed, which can be attributed to the change in tetragonality associated with the structural differences between the two models. Overall, these results indicate that the inclusion of long-range electrostatic interactions leads to moderate modifications in the structural and dielectric properties of BaTiO$_3$, slightly affecting the quantitative predictions while the overall qualitative trends remain consistent.

These findings indicate that the effect of long-range interactions on the PES is to modify its curvature and shift the positions of its energy minima, without altering the topological hierarchy of competing phases. Qualitative behaviors, which are governed by which phases are stable and how they connect, are therefore well captured by short-range MLPs, while quantitative properties such as transition temperatures, elastic constants, and dielectric constants, which are sensitive to the precise curvature and depth of the energy landscape, benefit from the inclusion of long-range interactions. This distinction provides a clear and practical criterion for selecting the appropriate level of MLP complexity depending on the target property.

## Supplementary Material

The supplementary material contains the comparison of energy and atomic forces prediction for MACE and MACELES, the size-dependent lattice constant change, and the polarization-electric field curve for $\varepsilon_a$.

## Acknowledgments

This study was supported by the Ministry of Education, Culture, Sports, Science and Technology (MEXT) (Nos. 24H00042), and New Energy and Industrial Technology Development Organization (NEDO). PYC would acknowledge the support of JST SPRING (Grant Number JPMJSP2108).

## Comflict of interest

None.

## Data avalability

The data that support the findings of this study are openly available in GitHub at https://github.com/nmdl-mizo/Finetuned-MACE-model.git.

# Supporting information

Long-range interaction effects on the phase transition, mechanical effect, and electric field response of BaTiO$_3$ by machine learning potentials

*Po-Yen Chen[1], Teruyasu Mizoguchi[1,2]*.

AUTHOR ADDRESS

[1]Department of Materials Engineering, the University of Tokyo, Tokyo, Japan

[2]Institute of Industrial Science, the University of Tokyo, Tokyo, Japan.

AUTHOR INFORMATION

**Corresponding Author**

poyen@iis.u-tokyo.ac.jp, teru@iis.u-tokyo.ac.jp

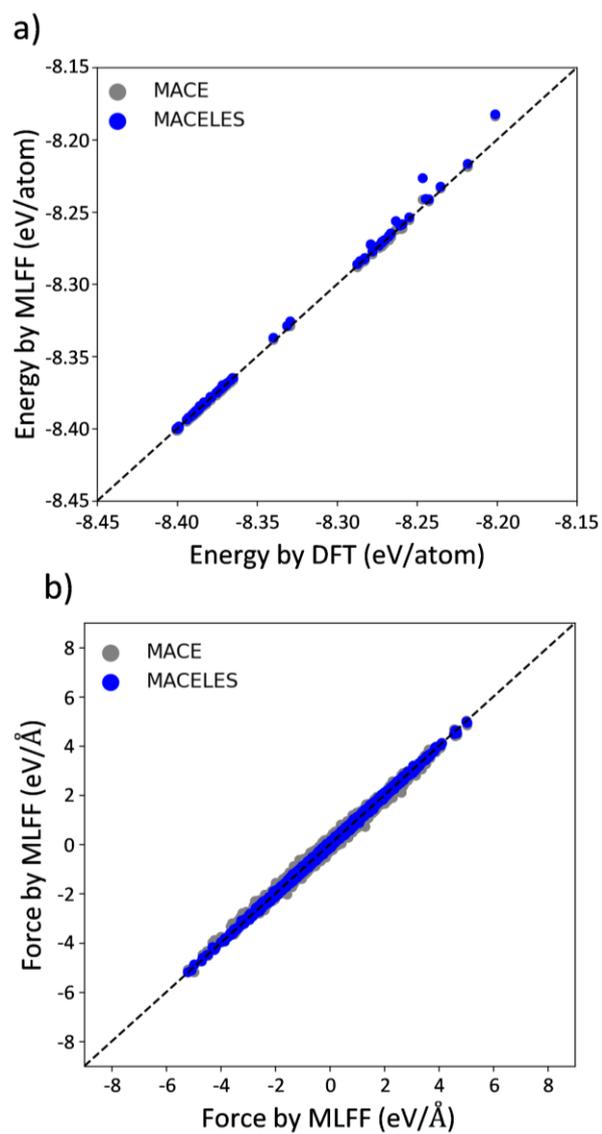

**Figure S1** The comparison of a) energy and b) atomic forces between machine learning potential and DFT for solid BaTiO3. The blue points show the prediction from MACELES, and the gray points are the prediction from MACE model.

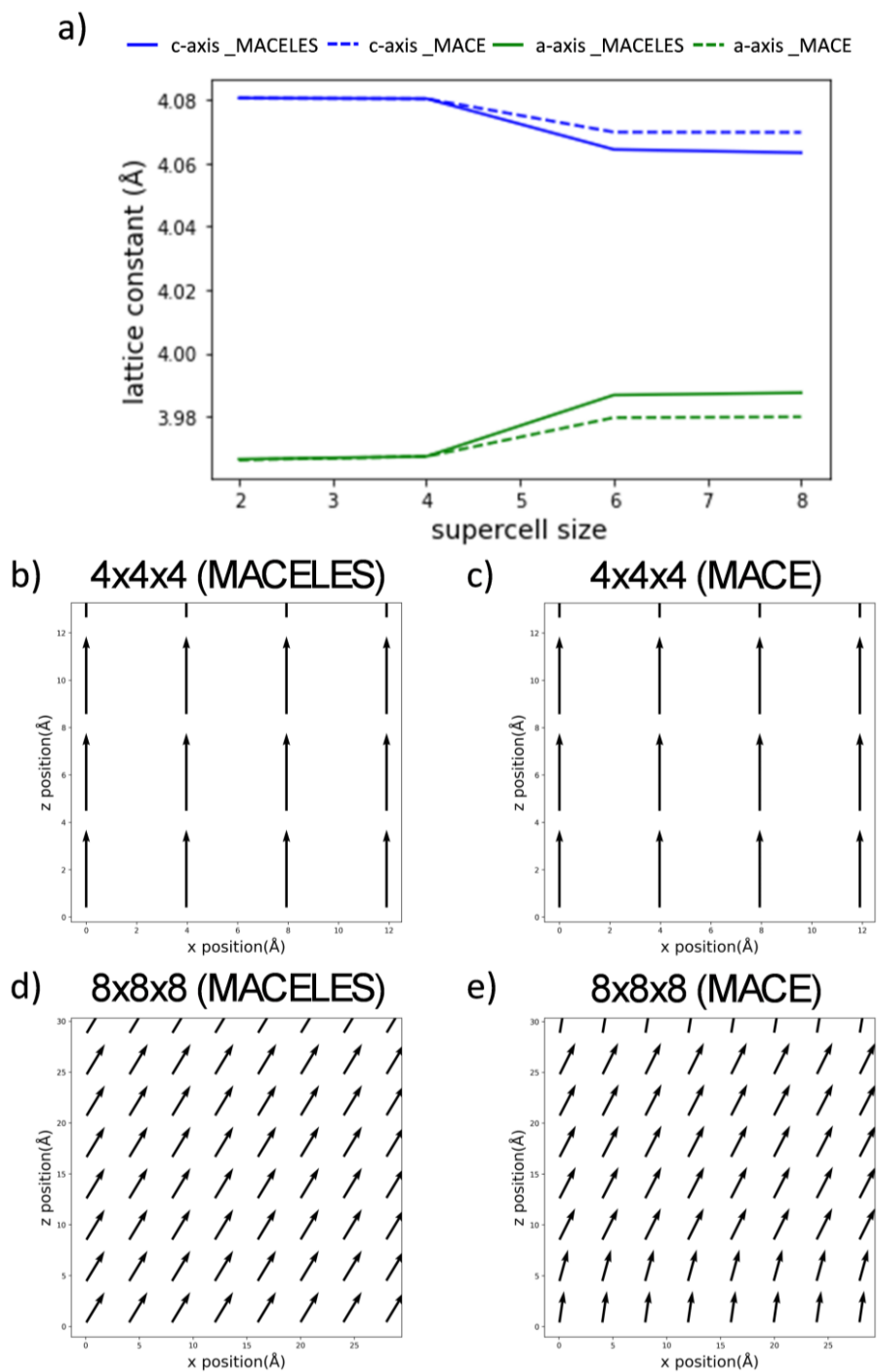

**Figure S2** a) the size-dependent lattice constant change, and the polarization orientation diagram for full relaxed 4x4x4 supercell by b) MACELES and c) MACE model and 8x8x8 supercell by d) MACELES and e) MACE model.

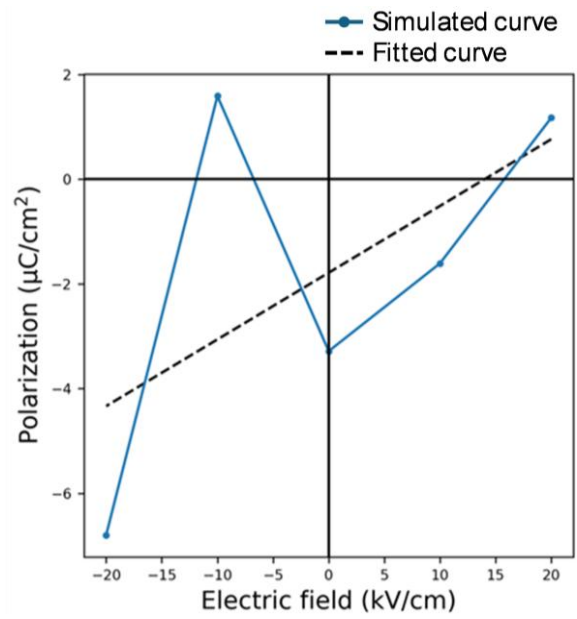

**Figure S3** The polarization-electric field curve for the linear response region for tetragonal BaTiO3 along a axis, the blue curve is the data from MD simulation, and the dashed line is the fitted curve.